\documentclass[twocolumn,showpacs,amsmath,amssymb]{revtex4}

\usepackage{graphicx}
\usepackage{bm}

\begin{document}

\title{Reflection of light from a moving mirror: \\
derivation of the relativistic Doppler formula
without Lorentz transformations}

\author{Malik Rakhmanov}

\affiliation{Department of Physics, University of Florida,
Gainesville, FL~32611}

\begin{abstract}
A special case of the relativistic Doppler effect, which occurs when 
light reflects from a moving mirror, is discussed. The classic formula 
for the Doppler shift is derived in a fully non-relativistic fashion 
using basic facts from geometry. The derivation does not involve
Lorentz transformations, length contractions and time dilations, and 
therefore is conceptually simpler than the standard derivations in 
physics textbooks. This discussion can be useful for teaching 
introductory physics and also for understanding the foundations of 
special relativity.
\end{abstract}

\pacs{03.30.+p, 03.65.Pm, 41.20.Jb}

%\keywords{Doppler effect, reflection of light}

\maketitle

The Doppler effect is commonly known as an apparent change in
frequency of a wave when the source of the wave or the observer are 
moving with respect to each other. Somewhat more complicated version
of this effect appears in special relativity in the context of light 
waves propagating in vacuum. In this case, the usual (nonrelativistic) 
Doppler contraction of the wavelength of light becomes mixed with the 
Lorentz contraction. For a source moving away from an observer with 
velocity $u$, special relativity predicts
\begin{equation}\label{relativity}
   \omega' = \sqrt{\frac{1 - u/c}{1 + u/c}} \;\; \omega ,
\end{equation}
where $\omega$ is the frequency of the wave measured in the rest-frame
of the source, and $\omega'$ is the frequency of the wave measured in
the rest-frame of an observer \cite{halliday}. This equation is usually 
derived in physics textbooks with the help of Lorentz transformations 
applied to the 4-dimensional wavevector \cite{jackson}. Sometimes, 
a derivation may not include Lorentz transformations explicitly, but
then it would rely on relativistic time dilations \cite{tipler}. 
Even more difficult is the derivation of the frequency change when
the wave reflects from a moving object (mirror). In this case, the 
answer is usually obtained by performing two Lorentz transformations: 
one from the laboratory frame to the rest frame of the mirror and 
the other in reverse. 

An alternative derivation can be obtained by noticing that the mirror 
forms an image which moves away from an observer. The observer then
detects the reflected wave as if it were coming from the image behind 
the mirror. In classical physics, a mirror moving with velocity $v$ 
creates an image moving with velocity $u = 2v$. In special relativity, 
the image would be moving with velocity
\begin{equation}\label{addVel}
   u = \frac{2v}{1 + v^2/c^2} ,
\end{equation}
which results from the law for relativistic velocity addition. 
Substituting Eq.~(\ref{addVel}) into Eq.~(\ref{relativity}), we 
obtain the formula for the frequency of the reflected light
\begin{equation}\label{doppler}
   \omega' = \frac{1 - v/c}{1 + v/c} \;\; \omega .
\end{equation}
Although this derivation seems simple enough, it still uses 
relativistic concepts. Namely, Eq.~(\ref{addVel}) is usually derived 
using Lorentz transformations.

It is interesting to note that the law for relativistic velocity
addition, of which Eq.~(\ref{addVel}) is a special case, can be derived 
directly from the constancy of the speed of light without any use of
Lorentz transformations \cite{mermin}. In this paper we show that even
the velocity addition formula is not necessary and give an entirely 
non-relativistic derivation of Eq.~(\ref{doppler}).

\begin{figure}[t]
   \centering\includegraphics[width=0.4\textwidth]{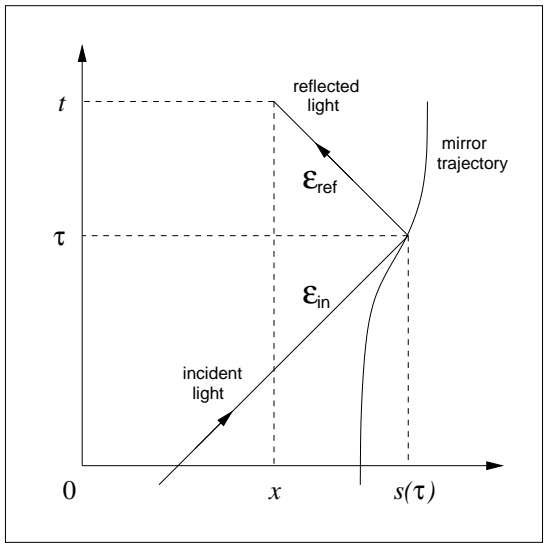}
   \caption{Light ray and mirror trajectory in $x-t$ space.}
   \label{xtspace}
\end{figure}

Consider an electromagnetic wave ${\mathcal{E}}(x,t)$ propagating 
in the positive $x$-direction and assume that the wave is incident upon 
a mirror which is moving along the trajectory: $x = s(t)$, as shown in 
Fig.~\ref{xtspace}. The mirror trajectory can be arbitrary provided
that the mirror velocity,
\begin{equation}\label{mirrVel}
   v(t) = \frac{d s}{d t} ,
\end{equation}
never exceeds the speed of light. The electric field measured by the 
observer at time $t$ must be the same as the field at the time of 
reflection $\tau$, when it coincides with input field:
\begin{equation}\label{in2ref}
   {\mathcal{E}}_{\mathrm{ref}}(x,t) =
      {\mathcal{E}}_{\mathrm{in}}[s(\tau), \tau] .
\end{equation}
The time of reflection can be found from the figure:
\begin{equation}\label{timeBounce}
   \tau = t - \frac{s(\tau) - x}{c} .
\end{equation}
This equation defines $\tau$ as an implicit function of $x$ and $t$, 
which means that in general we cannot solve for $\tau$. However, we
can find its derivatives with respect to $t$ and $x$:
\begin{equation}\label{dtau/dt}
    \frac{\partial \tau}{\partial t} = c \, 
    \frac{\partial \tau}{\partial x} = \frac{c}{c + v(\tau)} .
\end{equation}

For a plane-monochromatic wave with frequency $\omega$, the electric 
field is given by
\begin{equation}\label{monochrom}
   {\mathcal{E}}_{\mathrm{in}}(x,t) = \cos(\omega t - k x) ,
\end{equation}
where $k$ is the wavenumber: $k = \omega/c$. In this case, 
Eq.~(\ref{in2ref}) yields the reflected wave in the form:
\begin{equation}\label{E_ref}
   {\mathcal{E}}_{\mathrm{ref}}(x,t) =
      \cos \left[ \omega \tau - k s(\tau) \right] .
\end{equation}
Here the dependence of the electric field on $x$ and $t$ is hidden
in $\tau$. A different, but more familiar representation for the
electric field can be found by substituting $\tau$ from
Eq.~(\ref{timeBounce}) into Eq.~(\ref{E_ref}). The result can be
written as
\begin{equation}
   {\mathcal{E}}_{\mathrm{ref}}(x,t) =
      \cos \left[ \omega t + k x + \phi(x,t) \right] .
\end{equation}
The phase shift $\phi(x,t)$ depends on the mirror position at 
the time of reflection:
\begin{equation}
   \phi(x,t)  = - 2 k s(\tau).
\end{equation}

Once the wave is reflected by a moving mirror its frequency is no
longer constant; it depends on the position of the observer and the
time of the measurement. The instantaneous frequency of the reflected 
light is defined as a rate at which the total phase of the wave 
changes in time at a given point:
\begin{eqnarray}
   \omega'(x,t)
      & \equiv & \frac{\partial}{\partial t} \left[
                 \omega t + k x + \phi(x,t)  \right] \\
      & = & \omega + \frac{\partial \phi}{\partial t} .
\end{eqnarray}
Thus, the frequency of the reflected wave is shifted with respect to
the frequency of the incident wave by $\partial \phi/\partial t$. 
This partial derivative can be found using the chain rule: 
\begin{equation}\label{chainRule}
   \frac{\partial \phi}{\partial t} = - 2 k \; \frac{ds}{d\tau} \;
      \frac{\partial \tau}{\partial t}.
\end{equation}
The first derivative in the right-hand side of this equation,
$ds/d\tau$, is nothing but the mirror velocity at the time of reflection.
The second derivative is given by Eq.~(\ref{dtau/dt}). We thus find
the frequency of the reflected wave as
\begin{equation}\label{doppler2}
   \omega'(x,t) = \frac{c - v(\tau)}{c + v(\tau)} \;\; \omega ,
\end{equation}
\vspace{3mm}\noindent
which represents the relativistic Doppler effect and is an
extension of Eq.~(\ref{doppler}) to non-uniform mirror motions.

A natural question which one can ask is: what happens to the wavelength? 
The wavelength $\lambda$ can be found from the wavenumber, $\lambda=2\pi/k$, 
whereas the wavenumber is related to the frequency by
\begin{equation}
   \omega = c \, k .
\end{equation}
However, it is not clear that this relationship applies to the wave
reflected by the moving mirror because $\omega$ is no longer constant.
Furthermore, $k$ is not constant either. In this situation, the 
wavenumber shall be defined as a rate at which the total phase of the
wave changes in space, provided that time is frozen:
\begin{eqnarray}
   k'(x,t)
      & \equiv & \frac{\partial}{\partial x} \left[
                 \omega t + k x + \phi(x,t) \right] \\
      & = & k + \frac{\partial \phi}{\partial x} .
\end{eqnarray}
Expanding $\partial \phi/\partial x$ as in Eq.~(\ref{chainRule}), we
obtain
\begin{equation}
   k'(x,t) = \frac{c - v(\tau)}{c + v(\tau)}\;\; k ,
\end{equation}
which explicitly proves that 
\begin{equation}
   \omega'(x,t) = c \, k'(x,t) .
\end{equation}
Thus, the standard dispersion relation for electromagnetic waves in 
vacuum remains the same even if the waves are reflected from a mirror 
moving along an arbitrary trajectory.

\begin{acknowledgments}
The author would like thank N.D. Mermin for illuminating discussions.
This research was supported by the National Science Foundation 
under grant PHY-0070854.
\end{acknowledgments}

\bibliography{refs}

\end{document}